\documentclass[a4paper,fleqn,usenatbib]{mnras}
\usepackage{ae,aecompl}
\usepackage{graphicx}	
\usepackage{amsmath}	
\usepackage{amssymb}	
\usepackage{multicol}        
\usepackage{bm}		
\usepackage{pdflscape}	

\title[Environmental-dependence of AGN in SDSS]{The Dependence of AGN Activity on Environment in SDSS}

\author[Man et al.]{Zhong-yi Man,$^{1,2}$
Ying-jie Peng,$^{2}$\thanks{E-mail: yjpeng@pku.edu.cn}
Xu Kong,$^{3,4}$
Ke-xin Guo,$^{2,5}$
\newauthor Cheng-peng Zhang,$^{1,2}$
and Jing Dou$^{1,2}$\\
$^{1}$Department of Astronomy, School of Physics, Peking University, Beijing, 100871, China\\
$^{2}$Kavli Institute for Astronomy and Astrophysics (KIAA), Peking University, Beijing, 100871, China\\
$^{3}$Key Laboratory for Research in Galaxies and Cosmology, Department of Astronomy, \\
   University of Science and Technology of China, Hefei 230026, China\\
$^{4}$School of Astronomy and Space Sciences, University of Science and Technology of China, Hefei 230026, China\\
$^{5}$International Centre for Radio Astronomy Research, University of Western Australia, Crawley, WA 6009, Australia}

\date{Accepted 2019 June 17. Received in original form 2019 June 4}
\pubyear{2019}

\begin{document}
\label{firstpage}
\pagerange{\pageref{firstpage}--\pageref{lastpage}}
\maketitle

\begin{abstract}
Environment is one of the key external drivers of the galaxies, while active galactic nucleus (AGN) is one of the key internal drivers. Both of them play fundamental roles in regulating the formation and evolution of galaxies. We explore the interrelationship between environment and AGN in SDSS. At a given stellar mass, the specific star formation rate distribution of the AGN host galaxies remains unchanged with over-density, with the peak of the distribution around the Green Valley. We show that, at a given stellar mass, the AGN fraction that has been commonly used in previous studies (defined as the number of AGNs relative to all galaxies including passive and star forming ones) does decrease with increasing over-density for satellites. This is largely due to the fact that the fraction of passive galaxies strongly depends on environment. In order to investigate the intrinsic correlation between AGN and environment, especially under the assumption that AGN feedback is responsible for star formation quenching, the AGN fraction should be defined as the number of AGNs relative to the star-forming galaxies only. With the new definition, we find little dependence of AGN fraction on over-density, central/satellite, and group halo mass. There is only marginal evidence that AGN may prefer denser regions, which is possibly due to more frequent interaction of galaxies or higher merger rate in groups. Our results support the scenario that internal secular evolution is the predominant mechanism of triggering AGN activity, while external environment related processes only play a minor role.
\end{abstract}


\section{Introduction}

It is well established that observationally many galaxy properties, including star formation rate (SFR) and morphology, are strongly correlated with stellar mass and environment of the galaxies in both local Universe (e.g. \citealt{Kauf 03a, Kauf 04}; \citealt{Bal 06}; \citealt{Peng 10, Peng 12}; \citealt{Woo 13}; \citealt{Sch 14}), and to higher redshifts (\citealt{Dar 16}; \citealt{Kaw 17}). In the past decade, mass and environment have been considered as the most powerful drivers in quenching galaxies, namely, transforming star-forming galaxies into passive state (e.g. \citealt{Kauf 03b, Kauf 04, Bal 06, Peng 10}). 

AGN, powered by the growth of the supermassive black holes at the center of massive galaxies, is widely believed to play a fundamental role in regulating the star formation of its host galaxy and co-evolve with it (e.g. \citealt{Som 08}; \citealt{Kor 13}; \citealt{Hec 14}; \citealt{Cho 15}; \citealt{Har 17}). In fact, most galaxy formation and evolution models, including Semi-Analytic Models (SAMs) and hydro-dynamical simulations, usually require the feedback from AGN to produce many key observed features of the galaxy population, such as the stellar mass function (e.g. \citealt{Har 17} for a review). Although essentially all massive galaxies host a central supermassive black hole, not all galaxies become visible as AGN at a given epoch. Therefore, understanding the triggering mechanism of the AGN is one of the key issues in galaxy formation and evolution. 

AGN could be triggered by internal and/or external processes. It is well known that more massive galaxies are more likely to host AGN than low mass ones (\citealt{Dun 03}; \citealt{Flo 04}; \citealt{Bru 09}; \citealt{Pim 13}). Whether there is any dependence of AGN on the environment of its host galaxy is still under debate. A lot of literature suggests significant dependence of AGN on environment  (e.g. \citealt{Kauf 04, Gil 07, Best 07, Silver 09, Von 10, Bra 11, Hwa 12, Sab 13, Mar 13, Kha 14, Ehl 14, Silver 15, Col 17, Lop 17, Gor 18, Pow 18, Mag 18, Kou 18, Arg 18, Li 19, Kol 19}). For instance, \citet{Kauf 04} find that AGN host galaxies with strong [O III] emission are twice as frequent in low density regions than in high density regions. \citet{Gil 07} showed that X-ray selected AGNs at z $\sim$ 0 lie predominantly in areas of moderate dense regions. \citet{Silver 09} investigate X-ray selected AGN fraction as a function of galaxy overdensity using zCOSMOS spectroscopic survey data up to z $\sim$ 1. They find that massive galaxies harboring AGNs preferentially reside in lower-density regions, as is the case with studies of narrow-line AGN in SDSS. This is interpreted as that AGN activity requires a sufficient gas supply, which is likely to be adequate in underdense regions. \citet{Bra 11} investigate both X-ray and radio-loud AGNs within the UKIDSS Ultra-deep Survey in the redshift range from z $\sim$ 1.0 to 1.5 and find that both AGN types live in overdense environments. \citet{Ehl 14} found the fraction of X-ray bright AGNs increases with clustercentric distance. \citet{Lop 17} found that AGN favors in environments typical of the field, low mass groups or cluster outskirts, as the result of galaxy interactions. In SDSS DR7, low S/N LINERs are also found more likely to populate low density environments \citep{Col 17}. More recently, \citet{Kou 18} found X-ray selected AGNs in high mass clusters show opposite dependence on environment than in low mass clusters.   
  
Despite that many evidence support the significant environmental dependence of AGN, some studies claim the opposite. \citet{Mil 03} found that, in 4921 SDSS local galaxies, fraction of optical-selected AGN remains unchanged from the cores of galaxy clusters to the rarefied field population. Similarly, no significant difference in AGN fraction was observed between cluster and field galaxies (X-ray selected, \citealt{Mar 07}), or at different positions within the galaxy clusters (optical-selected, \citealt{Pim 13}). \citet{Nat 18} found the clustering of AGN is equivalent to that of galaxy, implying AGN has no preference for environment. \citet{Lix 18} found the clustering of narrow-line AGN with blue color or massive red galaxies is almost identical with controlled galaxies. In a most recent work, \citet{Ami 19} concluded that nuclear activity is weakly affected by local galaxy density using a sample including AGN and star-forming galaxies. 

Above apparently contradictory results on the environmental dependence of AGN could be attributed to different AGN selection criterion, the use of different environment tracers (see \citealt{Muld 12} for a detailed comparison), or even any potential observational bias. We argue a more important factor in exploring the interrelationship between environment and AGN is that the passive dead galaxies should not be included in the analysis, for two reasons. First, many passive dead galaxies were quenched at high redshift (e.g. $z>1$), especially massive ones. These galaxies could be quenched by AGN and/or stellar feedback  (or by any other physical mechanisms) at high redshift. They are not helpful for us to understand the interrelationship between AGN and environment at current epoch (i.e. at $z\sim0$). Second, many passive galaxies were quenched due to environmental effect such as strangulation (e.g. \citealt{Lar 80, Fel 10, Peng 15}). These galaxies were not quenched by AGN feedback and hence are not useful to explore the environmental dependence of AGN. Therefore, AGN fraction should be defined only with AGNs and star-forming galaxies to better reveal the intrinsic correlation between AGN and environment. Through this, we can eliminate the known strong correlation between passive fraction and environment (e.g. \citealt{Dre 80, Kauf 04, Bal 06, Peng 10, Peng 12}).

In this paper, we mainly focus on a local galaxy sample ($0.02\le z\le0.085$) in SDSS Data Release 7 (DR7, \citealt{Aba 09}) to study the environment-dependence of AGN activity. Details of our sample selection is given in section 2. The main results are in section 3, where we investigate the interrelationship between environment and AGN activity by introducing a modified definition of AGN fraction. We further discuss our results in section 4 and summary the main conclusions in section 5. We use a concordance $\Lambda$CDM cosmology model with $H_{0}$ = 70 km s$^{-1}$ Mpc$^{-1}$ , $\Omega_\Lambda$ = 0.75, and $\Omega_m$= 0.25 in this work. All magnitudes are quoted in the AB normalization.

\section{Data Set \label{Da}}
\subsection{The SDSS sample}
Our sample for analysis is based on the parent sample of \citet{Peng 10} drawn from SDSS DR7 \citep{Aba 09} catalog. The sample includes 238,474 objects with reliable spectroscopic redshift measurements in $0.02\le z\le 0.085$  (see \citealt{Peng 10} for details). Further, we retained galaxies only with well defined and reliable measurements of physical properties (e.g. stellar mass, SFR, metallicity), resulting in a sample containing 214,091 objects. These finally comprise the SDSS sample in this paper. 

Due to the SDSS spectroscopic selection $r<17.77$, the sample is incomplete below a stellar mass limit (e.g. 10$^{10.4}\, M_{\odot}$ at $z=0.085$). We therefore weight each galaxy below this limit by a factor of $1/V_{max}$ to correct the volume incompleteness within a given redshift bin ($V_{max}=1$ for galaxies above this limit). The $V_{max}$ values are derived from the \emph{k-correction} program v4\underline{~~}1\underline{~~}1 \citep{Bla 07}. The use of $V_{max}$ weighting allows us to include representatives of the galaxy population down to a stellar mass of about $10^9 M_{\odot}$. Besides, on average about 10\% of all SDSS targets are missed from the spectroscopy sample due to the minimum fiber spacing of 55 arcsec. To correct this, for each object, a spatial target sampling rate (TSR) is determined by using the fraction of objects that have spectra in the parent photometric sample within the minimum SDSS fiber spacing of 55 arcsec. In constructing the final population of SDSS galaxies used in our analysis, each galaxy is weighted by $1/V_{max}\times$1/TSR to correct both the volume incompleteness and local sampling rate.

Rest-frame absolute magnitudes are derived from the five SDSS $ugriz$ bands using the \emph{k-correction} program \citep{Bla 07} and further corrected onto the AB magnitude system. The stellar masses are determined from the same \emph{k-correction} code with \citet{Bruz 03} population synthesis models and a Chabrier IMF, which is consistent with that of \citet{Kauf 03a} and \citet{Gal 05} within $\sim$ 0.1 dex. 

\subsection{Overdensity}
We take overdensity $\delta$, the dimensionless density contrast, as an estimator of environment of individual SDSS galaxy: $\delta_i=(\rho_i-\rho_m)/\rho_m$. $\rho_i$ is the number density of galaxies within the projected radius to the 5$^{th}$ nearest neighbor (hereafter 5NN); $\rho_m$ is the (volume) mean density at a given redshift. Local density field of 5NN is calculated over a cylindrical volume with length of $\pm$1000 km s$^{-1}$ to exclude the galaxy proper motion effect. We use a volume-limited population of density tracers. In practice,  the spectroscopic sample with $M_{B,\, AB}\ \leq -19.3 - z$ are selected as density tracers. To avoid edge effects in local density measurement, we consider objects with $f > 0.9$, where $f$ is the fraction of the aperture adopted within the SDSS region \citep{Kov 10}.

\begin{table}
\begin{center}
\caption{Classifications of galaxies from the MPA-JHU DR7 release \citep{Bri 04}. In this work, star-forming galaxies refer to the first subsample which is about 40\% of the full sample. \label{tbl-1}}
\begin{tabular}{crrrrrrrrrr}
\hline\hline
Subsample & Number & Percentage \\
\hline
Star-forming & 85528 & 39.95\% \\
Low S/N Star-forming &39853 &18.61\% \\
Composite &16428 & 7.67\%  \\
AGN &8005 &3.74\%\\
Low S/N LINER & 19277 &9.00\%\\
Unclassifiable & 44701 & 20.88 \%\\
\hline
\end{tabular}
\end{center}
\end{table}

\subsection{Group Catalog and Halo Mass}

Our galaxy group catalog is an updated version of \citet{Yang 07} catalog. It contains galaxies in the Main Galaxy Sample of NYU Value-Added Galaxy Catalog (NYU-VAGC, \citealt{Bla 05, Ade 08, Pad 08}) within $0.01\leq z \leq 0.20$. By implementing their group finder, \citet{Yang 07} divide galaxies into groups in which each galaxy is marked as central or satellite. The dark matter halo mass $M_h$ of each group is estimated based on the intergrated stellar mass or luminosity of all group members above a certain luminosity limit. In principle, the group with highest total stellar mass (or luminosity) is assigned the highest halo mass following a given halo mass function (HMF). In this work, we will use the halo mass derived from total stellar mass because it is less affected by ongoing star formation.
  
\begin{figure}  

\includegraphics[width=1\columnwidth]{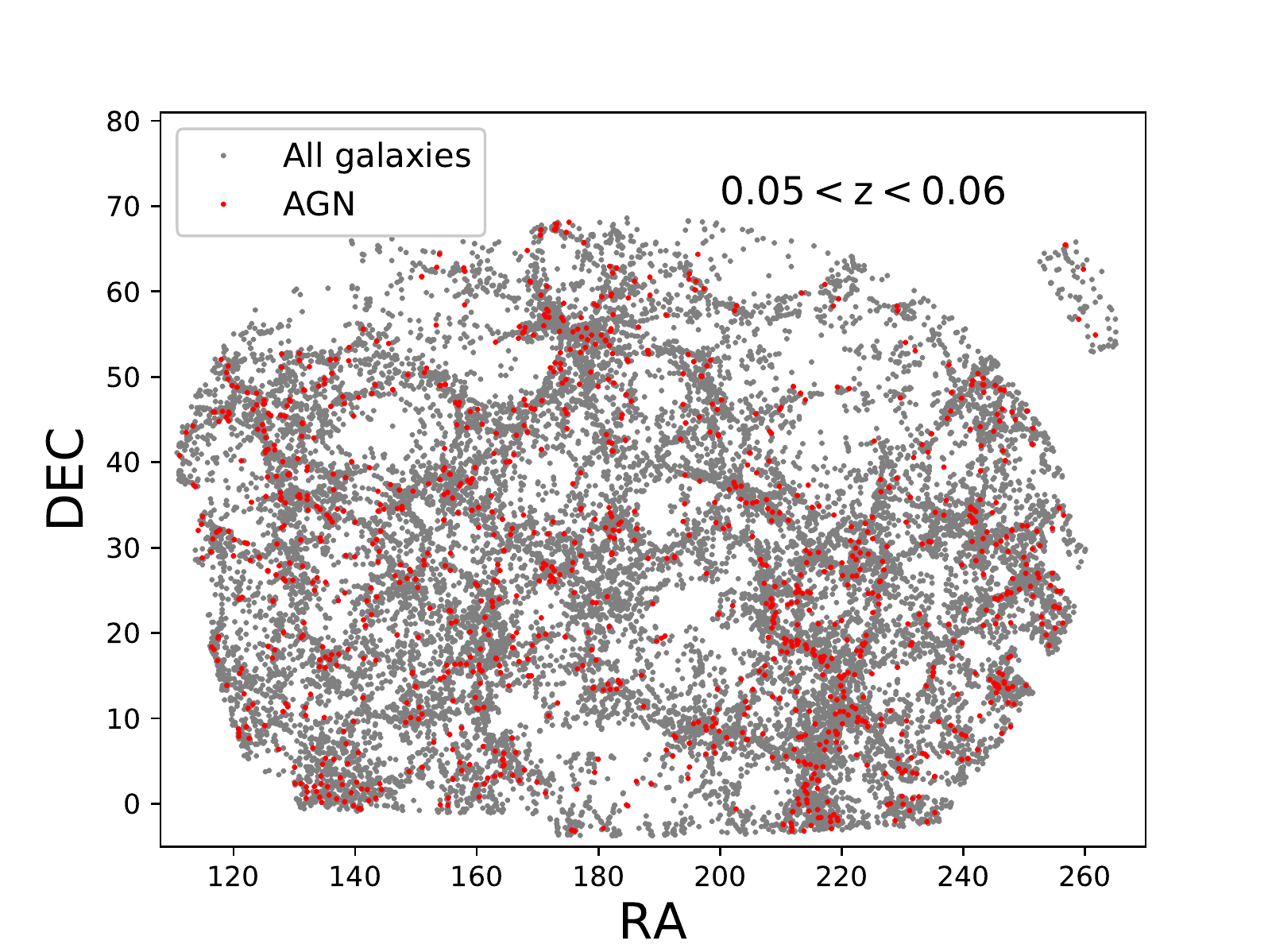}
\caption{Spatial distribution of SDSS galaxies in $0.05<z<0.06$. AGN host galaxies (red points) are plotted overlapping on all galaxies (grey points). One can see the prominent features of large-scale structures: filaments, nodes, voids and clusters. AGN distribution basically follows the large-scale structure of all galaxies by a rough visual inspection.\label{figure 1}}
\end{figure}

\subsection{SFR Estimates and Classifications}
The SFR and types of galaxies are directly retrieved from the MPA-JHU DR7 release \footnote{https://wwwmpa.mpa-garching.mpg.de/SDSS/DR7/}, where galaxies are further classified into star-forming, composites and AGNs \citep{Bri 04} based on the BPT diagram \citep{BPT 81}. The star-forming population here refer to those with S/N $>3$ in all four BPT lines, excluding the low S/N star-forming galaxies with only S/N $>$ 2 in H$_\alpha$. Composites are defined as galaxies somewhat in between AGN and star-forming on the BPT diagram. Low signal-to-noise ratio (S/N) LINERs refer to galaxies with [N II]/H$_\alpha>0.6$ and S/N >0.3 in both lines, but too low S/N for [O III] and H$_\beta$. Unclassifiable galaxies have no or very weak emission lines hence unable to be classified by BPT. The number and percentage of each category is listed in Table \ref{tbl-1}.

The SFR of star-forming galaxies are derived based on their H$\alpha$ emission line luminosities, corrected for extinction using the H$\alpha$/H$\beta$ ratio; the SFR of composites, low S/N LINERs and AGN host galaxies are estimated based on a likelihood distribution for specific star formation rate ($\rm sSFR=SFR/M_*$) as a function of D4000 constructed using the star-forming sample \citep{Bri 04}. Aperture effects are further corrected for all galaxies. The SFR is converted from Kroupa to Chabrier IMF using log\ SFR (Chabrier) = log\ SFR (Kroupa) $-$ 0.04.

\section{Results \label{result}}
\subsection{Environment}

\begin{figure*}
\centering
\includegraphics[width=0.8\paperwidth]{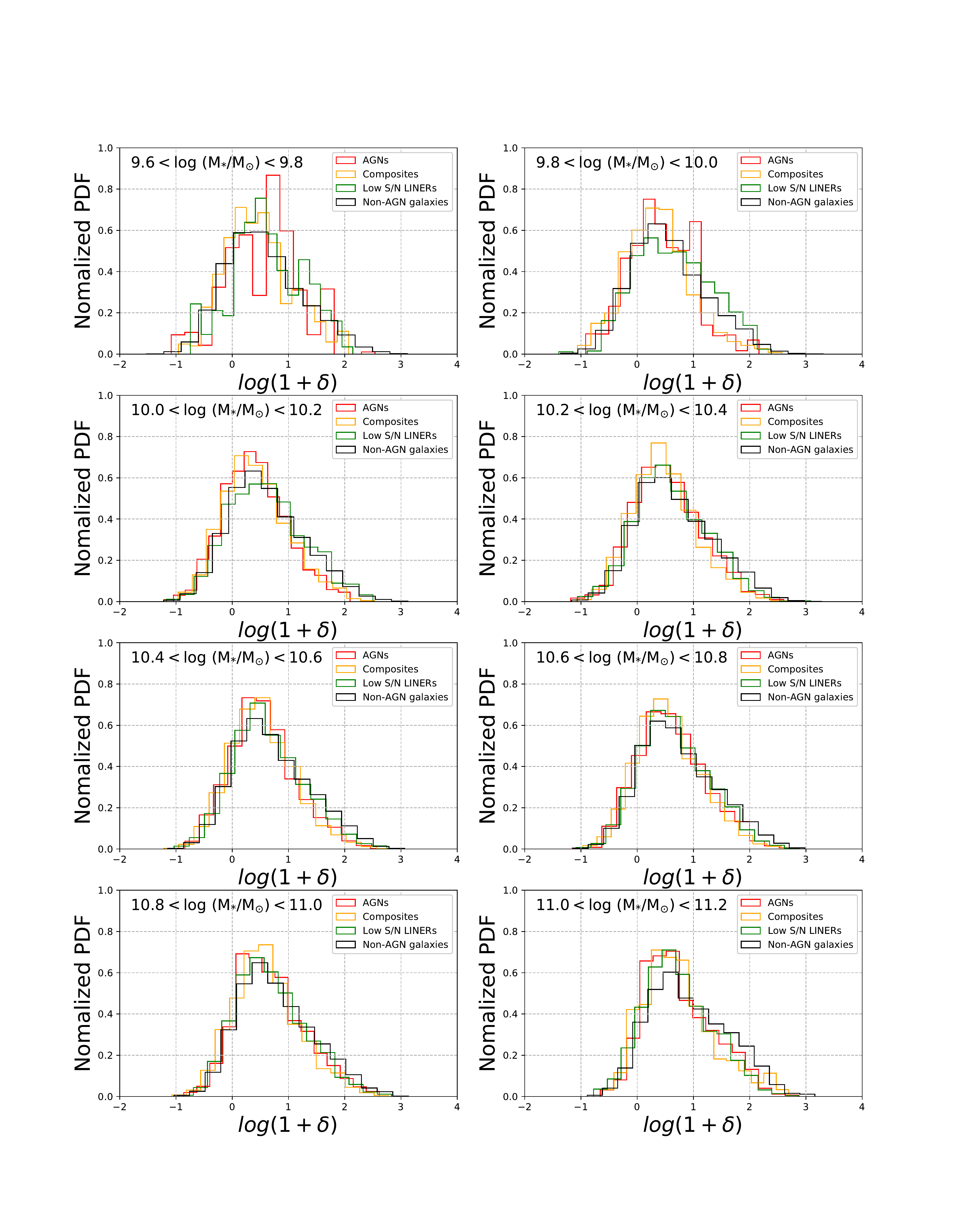}
\caption{The overdensity distributions for AGNs, composites, low S/N LINERs and non-AGN galaxies at different stellar masses. In general, there are no significant variance in their overdensity distributions. There may only be a slight difference that AGNs and composites have smaller high-density tails than low S/N LINERs and non-AGN galaxies. \label{figure 2}} 
\end{figure*}

\begin{figure*}
\includegraphics[width=0.8\paperwidth]{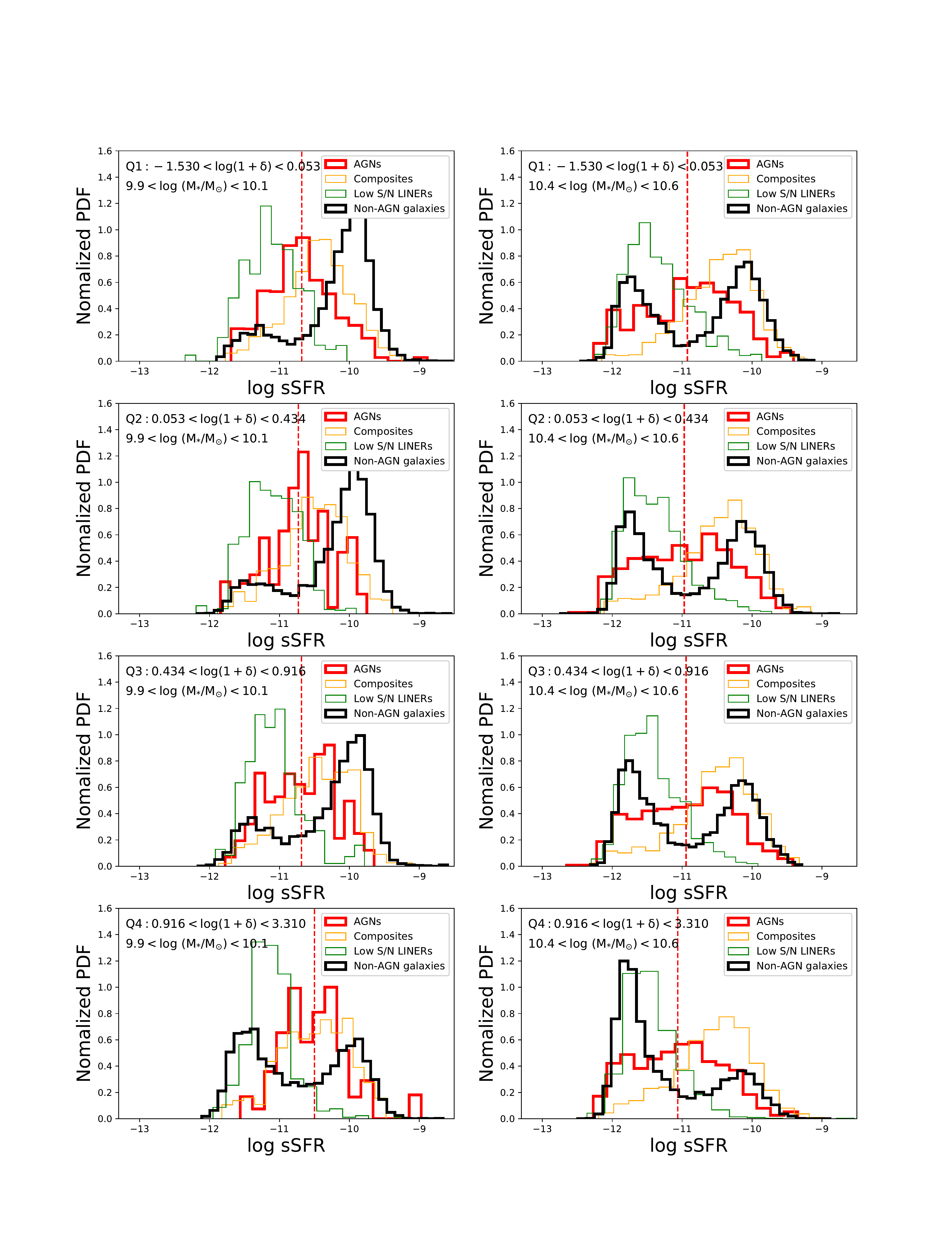}
\caption{The sSFR distributions for AGNs (in bold), composites, low S/N LINERs and non-AGN galaxies (in bold) at two fixed stellar mass $\rm 10^{10}\, M_{\odot}$ (first column) and $\rm 10^{10.5}\, M_{\odot}$ (second column). The galaxies are divided into overdensity quarters (Q1 for row 1, etc.) with equal amount in each quarter. The red dashed line in each panel represents the mean value of sSFR of AGN host galaxies. As overdensity increases, the star-forming peak of non-AGN host galaxies drops while the passive peak rises. AGNs and composites change little with overdensity, and are located in between the star-forming and passive populations. \label{figure 3}}
\end{figure*}

The most direct way to study the local environment of galaxies is to see how they are distributed in space. Here, we simply cut a redshift slice at $0.05\le z\le 0.06$ (for no particular reason) to provide a 2D view of the spatial distribution of AGNs and galaxies. In Figure \ref{figure 1}, all galaxies are plotted in grey dots, while AGN host galaxies are plotted in red dots. One can see the prominent features of large-scale structures: filaments, nodes, voids and several large clusters. AGNs seem to distribute following all galaxies. Although a large number of AGNs are relatively concentrated in galaxy clusters, there are still some AGNs in filaments or even in field. AGN distribution basically follows the large-scale structure of all galaxies and it is hard to find significant environmental dependence of AGN by a rough visual inspection. This motivates us to investigate the interrelationship between AGN and environment in more details.

There are various approaches to define local environment e.g. by distance to cluster center, group identity (central or satellite), location (cluster/filament/field), overdensity, dark matter halo mass and so forth. Overdensity $\delta$, for example, has been used as an efficient indicator of local environment which can be easily defined in observation (e.g. \citealt{Bol 10, Mor 10, Peng 12, Jian 12, Kov 14, Dar 16}). As argued in \citet{Muld 12}, the local environment which corresponds to scales internal to a halo is best probed using the nearest-neighbour methods. We hence use the overdensity defined by the 5NN as discussed in section 2.2.

In Figure \ref{figure 2}, we present the normalized probability distribution function (PDF) of overdensity for different types of galaxies. As we know, AGNs are tightly related to the stellar mass of their host galaxies: more massive galaxies are more likely to host AGNs (\citealt{Dun 03}; \citealt{Flo 04}). We therefore bin galaxies in stellar mass to exclude the mass-related effect. In general, AGNs, composites, low S/N LINERs and non-AGN galaxies basically share the same overdensity distribution. The only difference may be that in $\rm 9.8 < log (M_*/M_{\odot}) < 10.2$, AGNs and composites tend to have smaller high-density tails than low S/N LINERs and non-AGN galaxies. This could be due to an excess of passive galaxies in low S/N LINERs and non-AGN host galaxies as a result of environmental quenching.

 \begin{figure*}
\centering
\includegraphics[width=1\columnwidth]{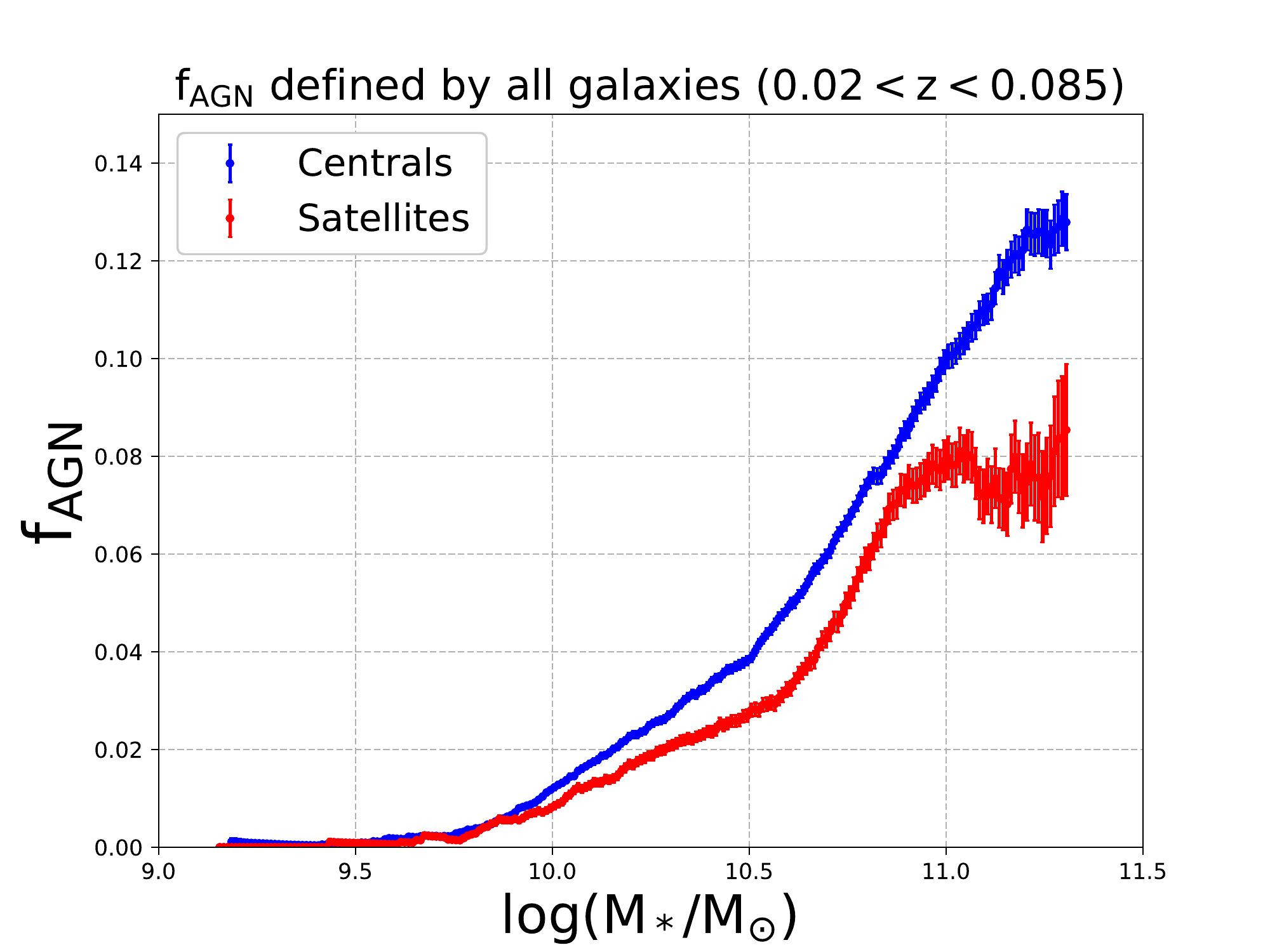}
\centering
\includegraphics[width=1\columnwidth]{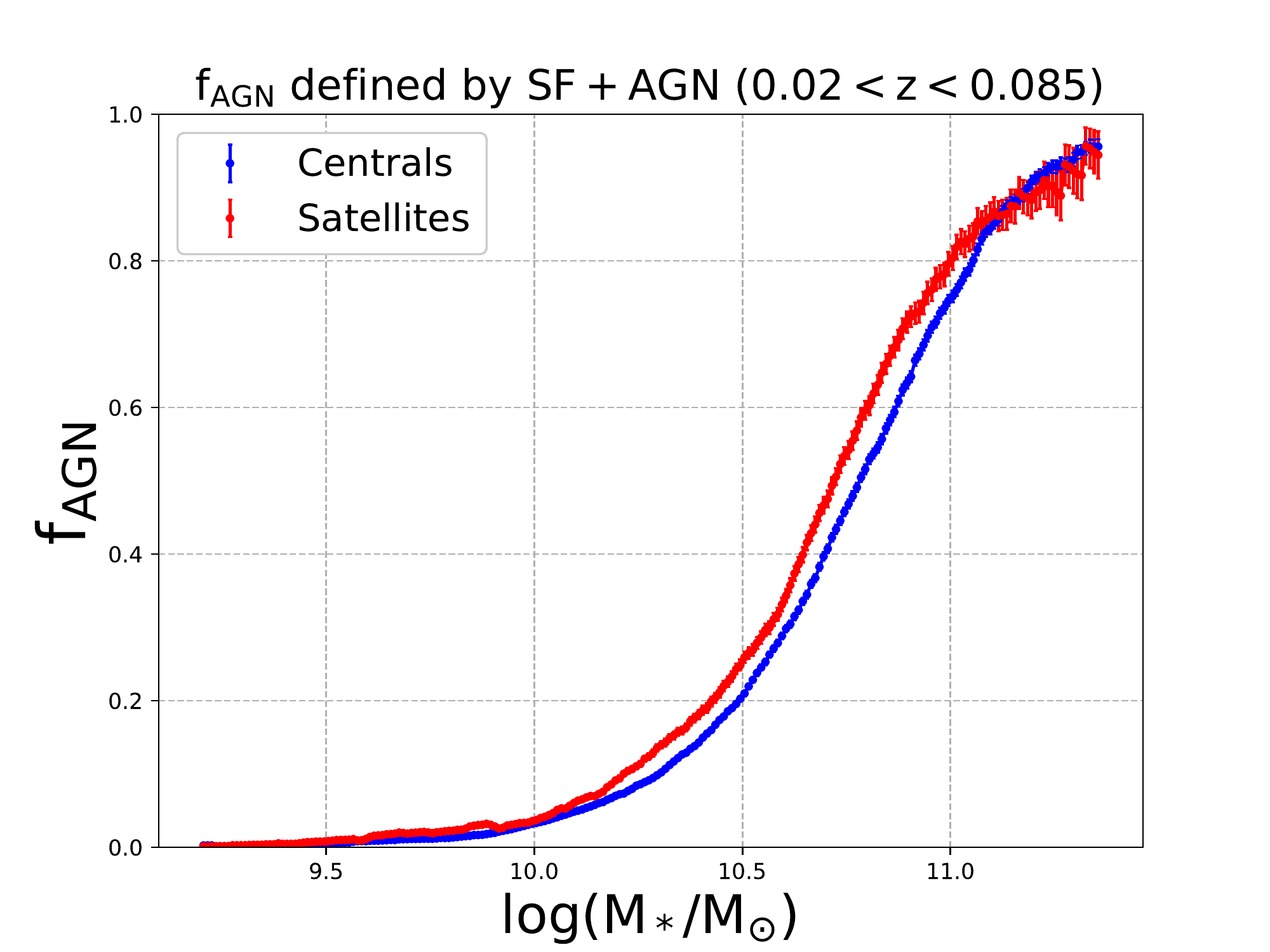}
\centering
\includegraphics[width=1.\columnwidth]{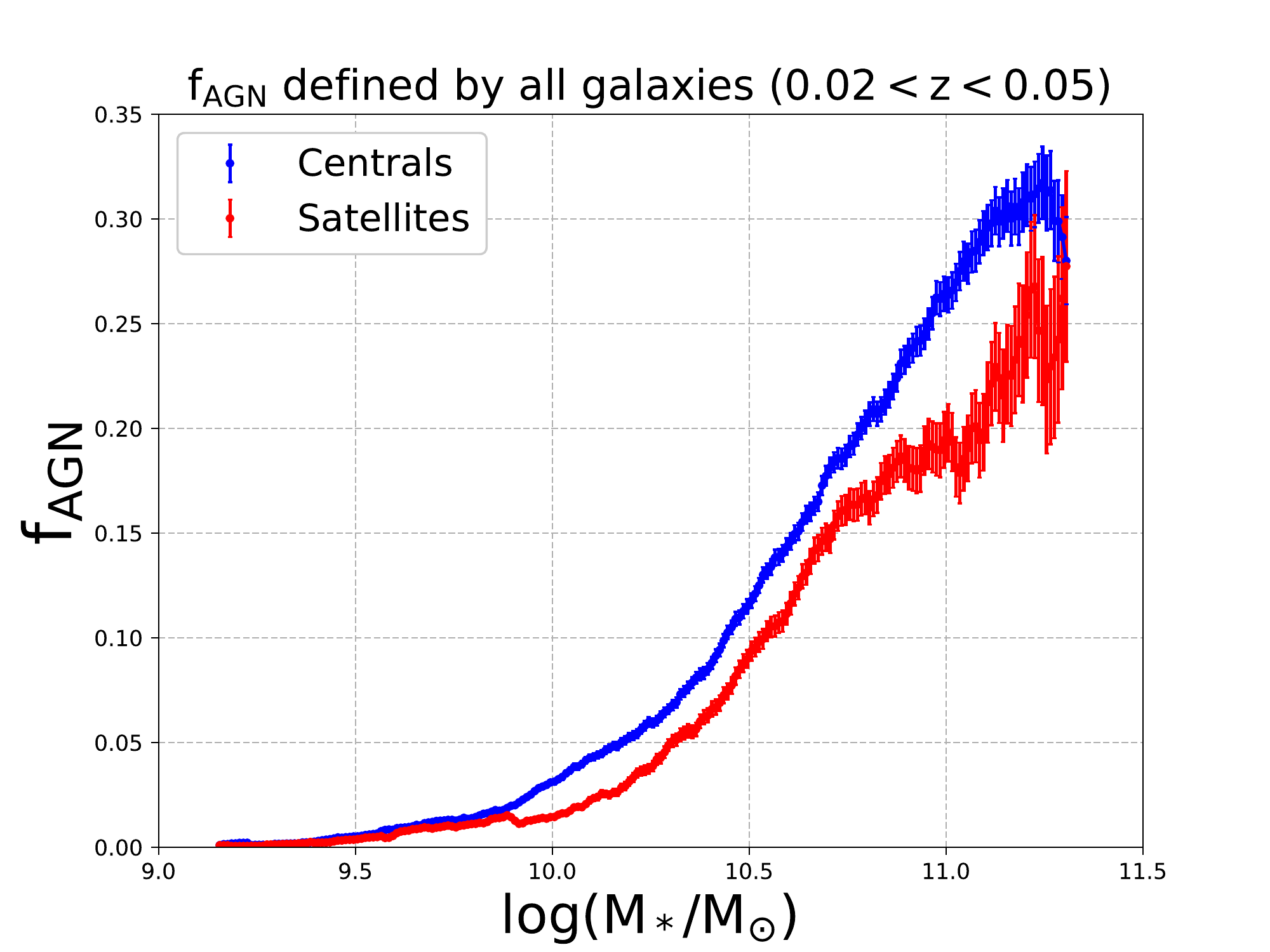}
\centering
\includegraphics[width=1\columnwidth]{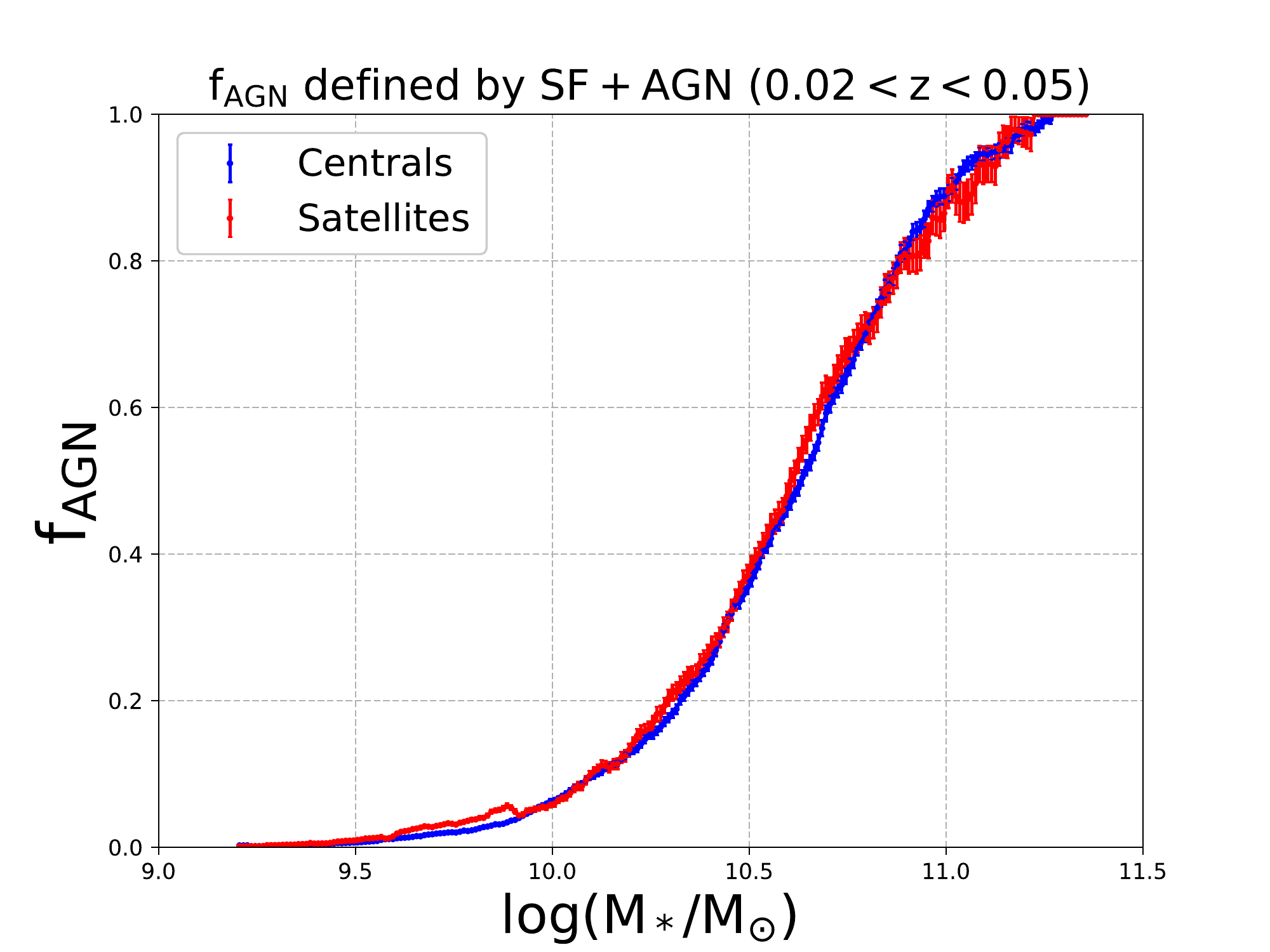}
\caption{AGN fraction as a function of stellar mass for central and satellite galaxies in $0.02<z<0.085$ (top panels) and $0.02<z<0.05$ (bottom panels). The fraction is defined by all galaxies (left), or by star-forming galaxies+AGN (right). Error bars are 1$\sigma$ of binomial distributions. In general, the AGN fraction increases with the stellar mass of the host galaxies. By using the AGN fraction defined by all galaxies, we find central galaxies on average have a higher AGN fraction than satellite galaxies. By using the AGN fraction defined by SF+AGN, we find the difference between centrals and satellites is largely reduced, or even disappears in the lower redshift range ($0.02<z<0.05$) where the stellar mass are more complete. \label{figure 4}}
\end{figure*}

AGN is believed to play an important role in regulating the star formation of its host galaxy, it would be hence helpful to examine the sSFR distribution of AGN hosts at different environments. In Figure \ref{figure 3}, we compare the normalized PDF of sSFR among AGN hosts, composites and non-AGN galaxies by quartering them in terms of overdensity $\delta$ at two fixed stellar mass $\rm 10^{10}\, M_{\odot}$ and $\rm 10^{10.5}\, M_{\odot}$, respectively. The distributions of AGNs and non-AGN galaxies are highlighted in bold. It is clear that the distribution of non-AGN host galaxies varies with overdensity: as overdensity increases, the star-forming peak drops while the passive peak rises, which is a result of the more violent environmental quenching in higher density regions. AGN hosts and composites are found mostly located in the ``green valley'', the transition area between the star-forming and passive populations. This indicates AGN activities may be associated with the quenching process of their host galaxies, which has been demonstrated in a number of studies (e.g. \citealt{Sal 07, Mul 12, Har 13, Gur 15, Les 16}). Moreover, one would expect the sSFR distribution of AGN hosts changes with density, if a strong environment dependence of AGN activity is assumed. However, it is shown that the distribution of AGN hosts and composites change little with environment, providing no evidence for a tight correlation between AGN and environment. Yet there is a caveat here that the SFR of AGN hosts and non-AGN host galaxies are derived in different ways, which might lead to some systematic difference.

\begin{figure*}
\centering      
\includegraphics[width=1\columnwidth]{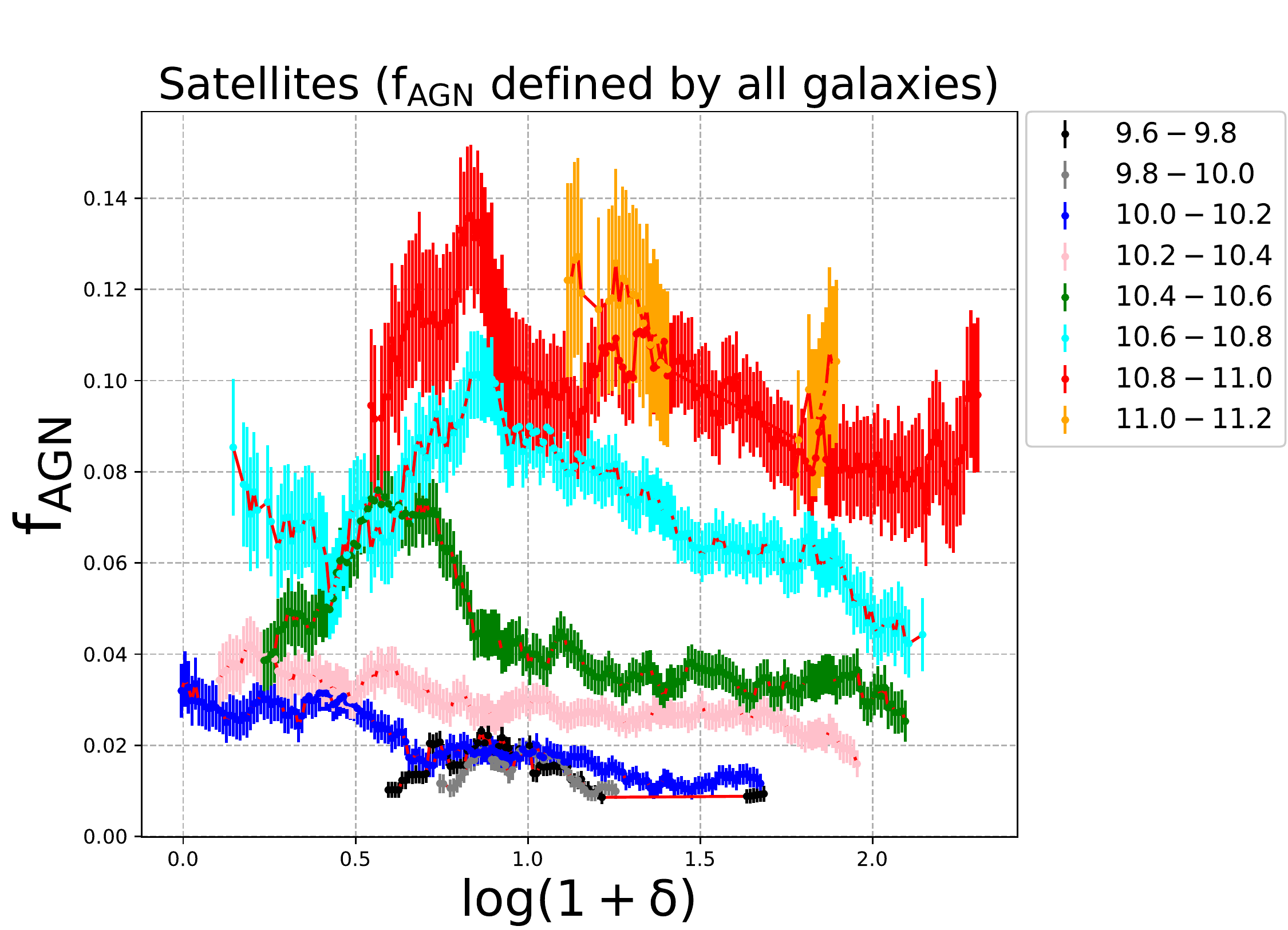}
\centering
\includegraphics[width=1\columnwidth]{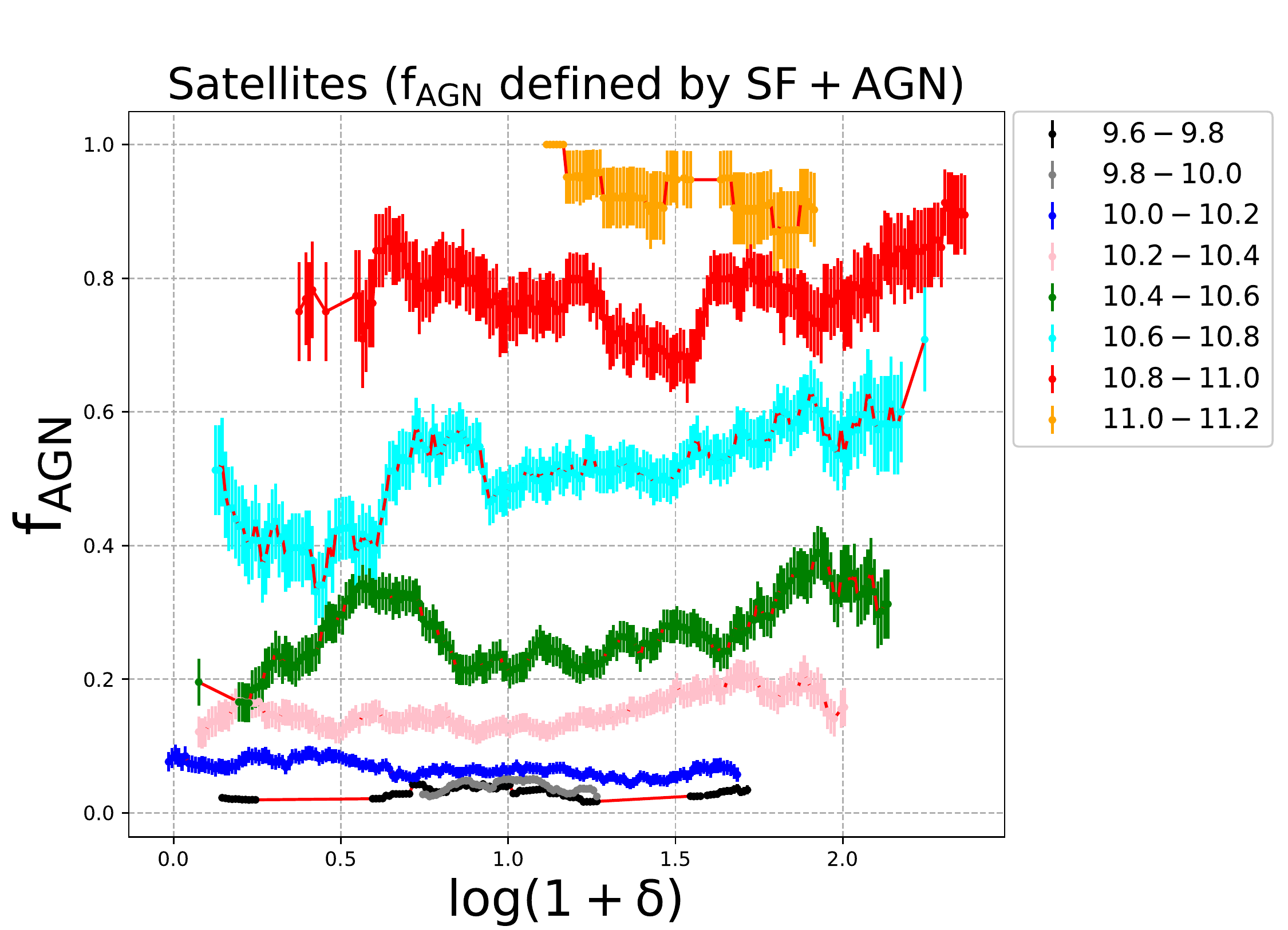}
\centering
\includegraphics[width=1\columnwidth]{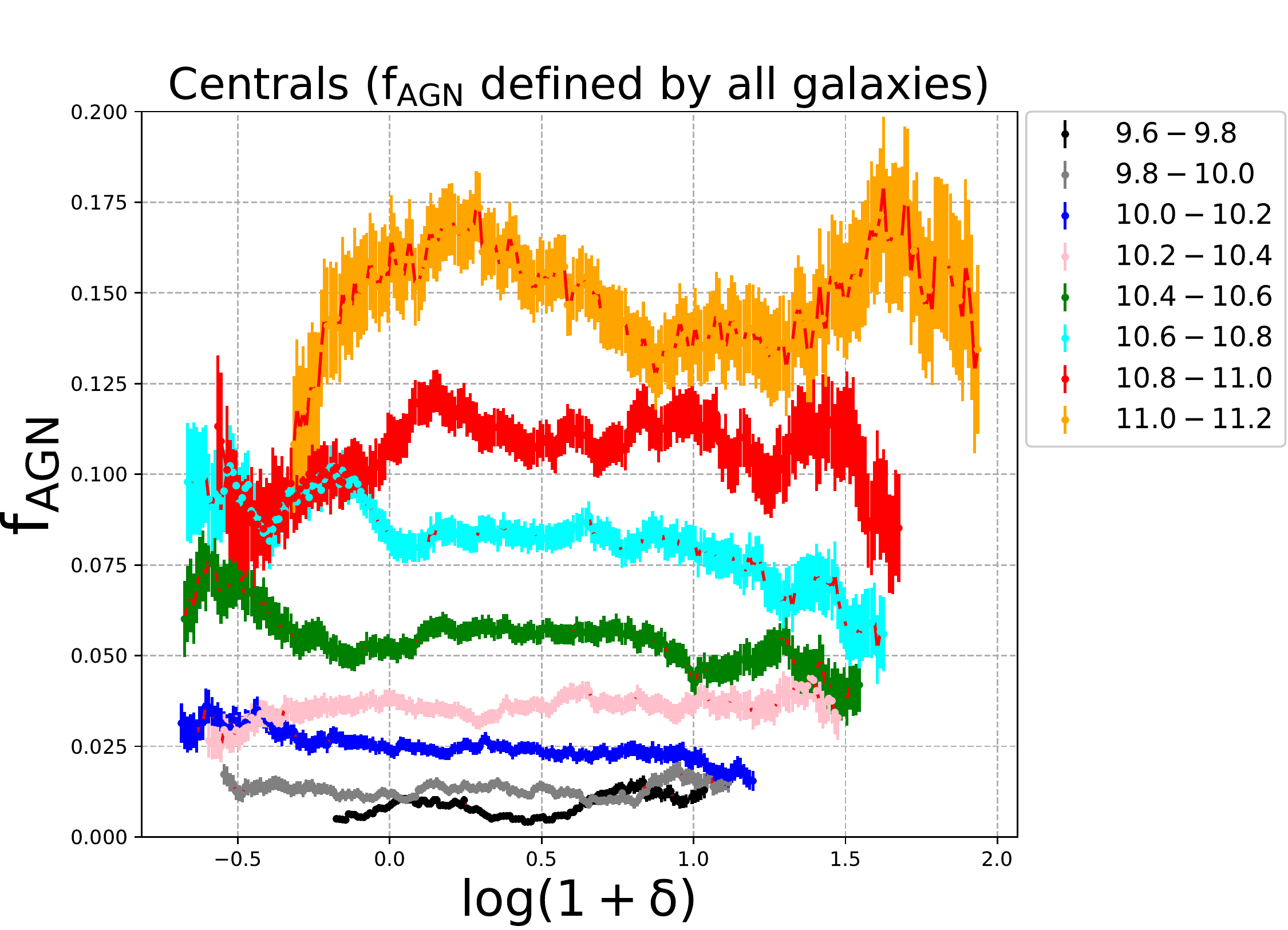}
\centering
\includegraphics[width=1\columnwidth]{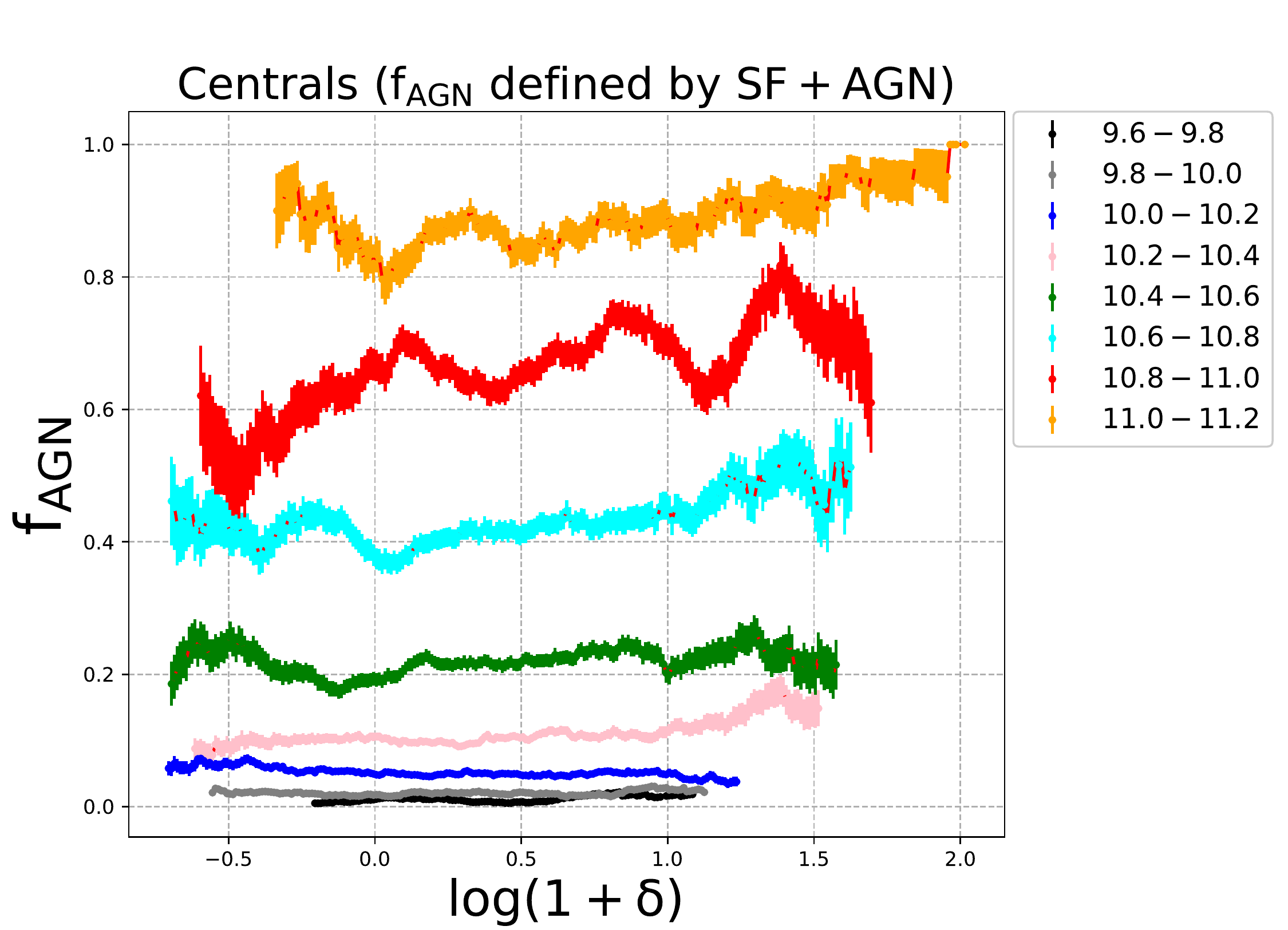}
\caption{AGN fraction as a function of overdensity for satellite galaxies (top panels) and central galaxies (bottom panels). The fraction is defined by all galaxies (left) and by star-forming galaxies+AGN (right). Error bars in different color represent different bins in $\rm log(M_*/M_{\odot})$. For satellite galaxies, we find the AGN fraction defined by all galaxies decrease significantly with overdensity, while the AGN fraction defined by SF+AGN changes little or only mildly increases with overdensity. For central galaxies, the AGN fraction does not change significantly with overdensity under both definitions. \label{figure 5}}
\centering
\includegraphics[width=1.2\columnwidth]{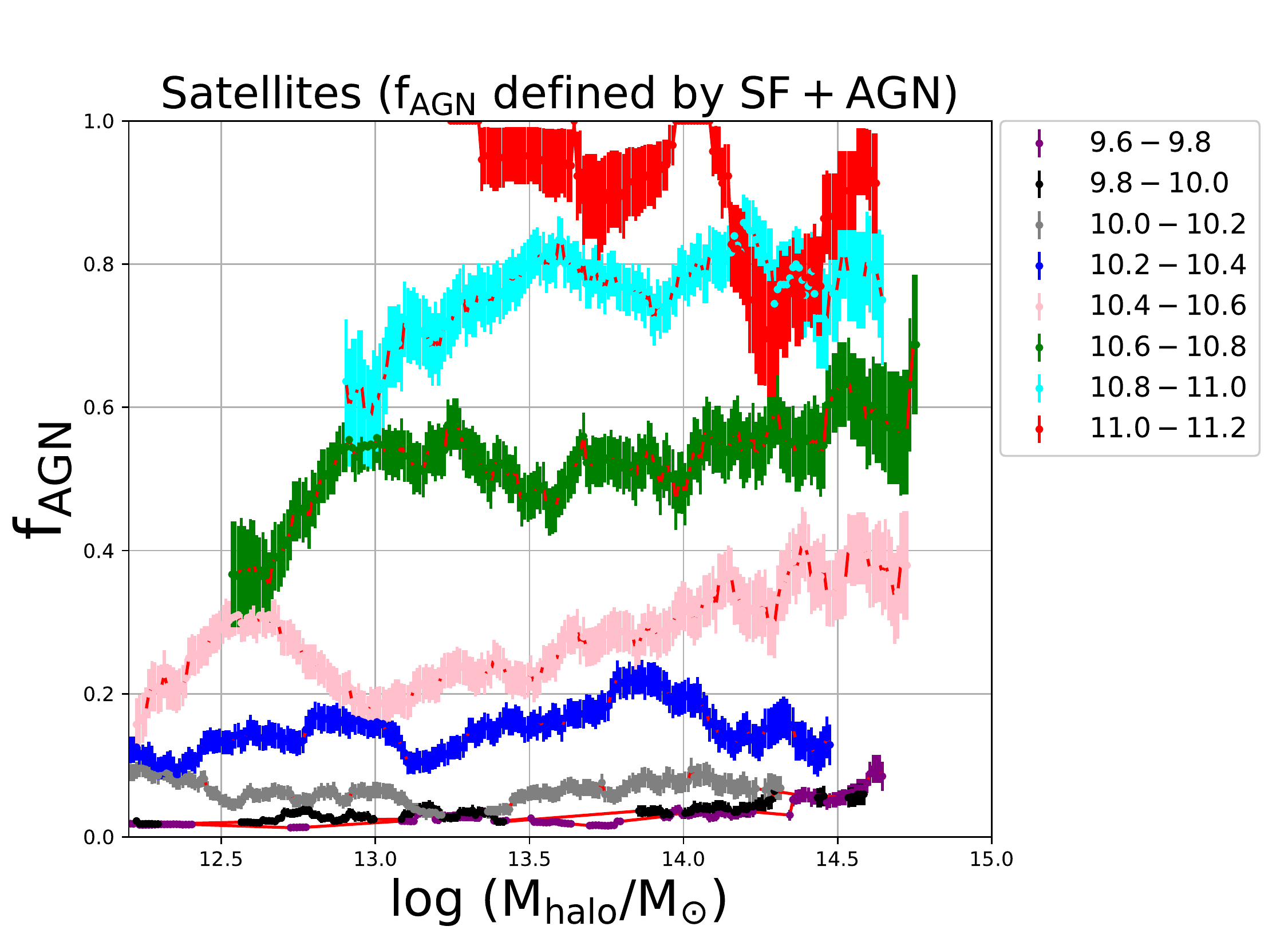}
\caption{AGN fraction (defined by star-forming galaxies+AGN) as a function of group halo mass for satellite galaxies. Error bars in different color represent different bins in $\rm log(M_*/M_{\odot})$. By using the AGN fraction defined by star-forming galaxies+AGN, we find the AGN fraction of satellites changes little or only mildly increases with overdensity. \label{figure 6}}
\end{figure*}

\subsection{AGN Fraction}  
A practical way to investigate the environmental dependence of AGN activity is to directly measure how frequently AGN activity is triggered in galaxies with different environments. Usually, the occurrence of AGN is quantified by AGN fraction defined as the number of AGNs relative to all galaxies including passive and star forming ones (e.g. \citealt{Silver 09}; \citealt{Hwa 12}; \citealt{Mar 13}):

\begin{equation}
\label{eq_all}
\rm f_{AGN}=\frac{N_{AGN}}{N_{All}}
\end{equation}
where $\rm N_{AGN}$ is the number of AGN host galaxies, $\rm N_{All}$ is the number of all galaxies

By using this definition, we examined the AGN fraction as a function of stellar mass for central and satellite galaxies, respectively. As shown in the top-left panel of Figure \ref{figure 4}, AGN fraction increases with stellar mass with a significant dichotomy between central and satellite: central galaxies are more likely to host AGN than satellite galaxies at a fixed stellar mass, especially for massive galaxies with $\rm log\ (M_*/M_{\odot})>11.0$. 

To better investigate the intrinsic correlation between AGN and environment, we propose a modified version of AGN fraction defined as the number of AGNs relatively to the star-forming galaxies only (which we will discuss in detail in section \ref{D}):

\begin{equation}
\label{eq_sf}
\rm f_{AGN}=\frac{N_{AGN}}{N_{AGN}+N_{SF}}
\end{equation}
where $\rm N_{AGN}$ is the number of AGN host galaxies, $\rm N_{SF}$ is the number of star-forming galaxies

The top-right panel of Figure \ref{figure 4} shows the AGN fraction as a function of stellar mass by using our new definition. The difference between centrals and satellites is largely reduced, and even disappears at low mass ($\rm log\ (M_*/M_{\odot})<10.1$) and high mass ($\rm log\ (M_*/M_{\odot})>11.0$) ends. In particular, if we narrow the redshift range down to $0.02 < z < 0.05$ where the stellar mass are more complete, we will find basically little difference between centrals and satellites (see the bottom panels of Figure \ref{figure 4}), while the dichotomy remains still distinct by using the previous AGN fraction defined by all galaxies. We can hence say that by using the new AGN fraction, AGN activity manifests no significant dependence on the local environment of its host galaxy in terms of central/satellite dichotomy. 

\begin{figure*}
\centering
\includegraphics[width=1\columnwidth]{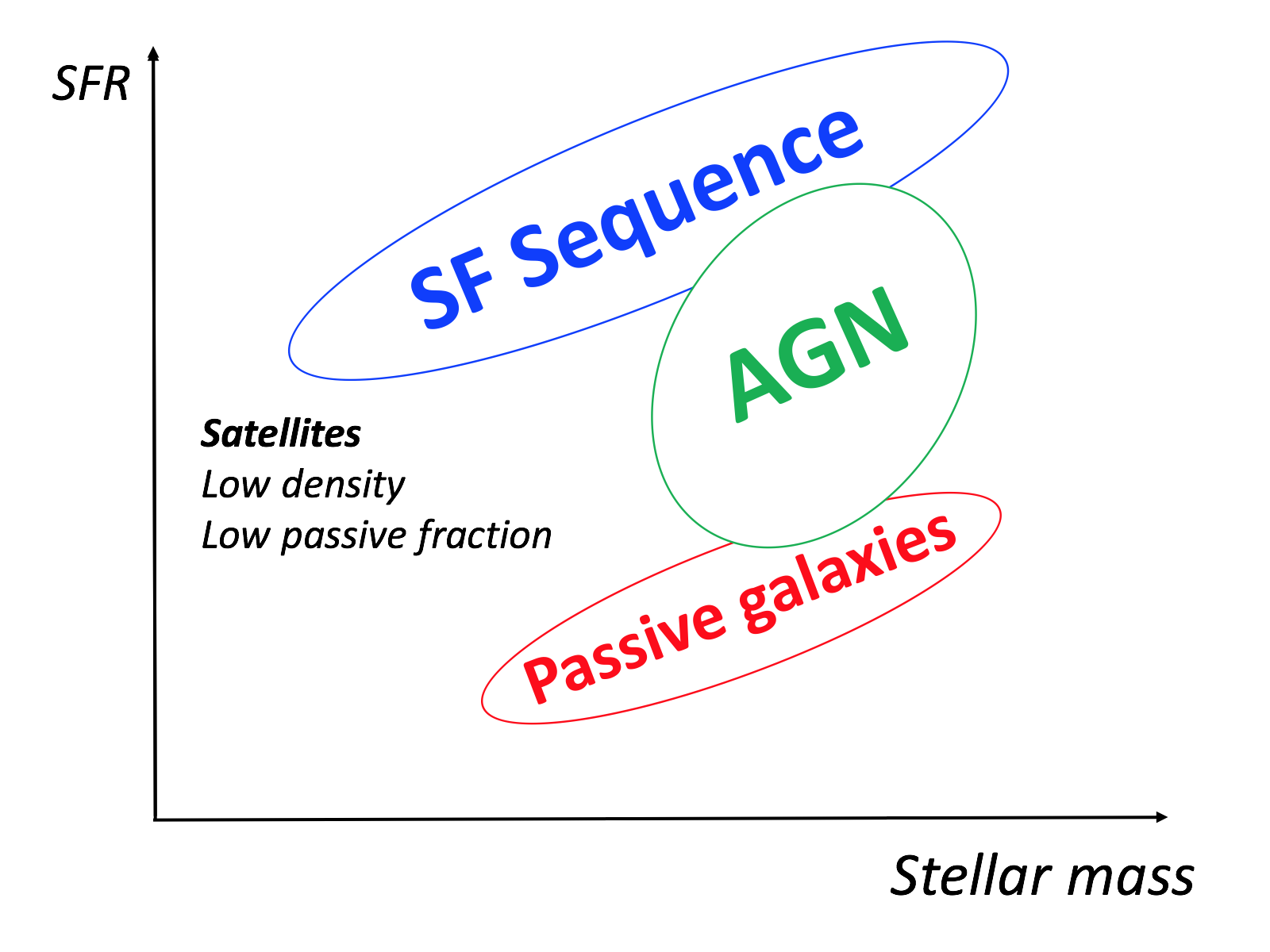}
\centering        
\includegraphics[width=1\columnwidth]{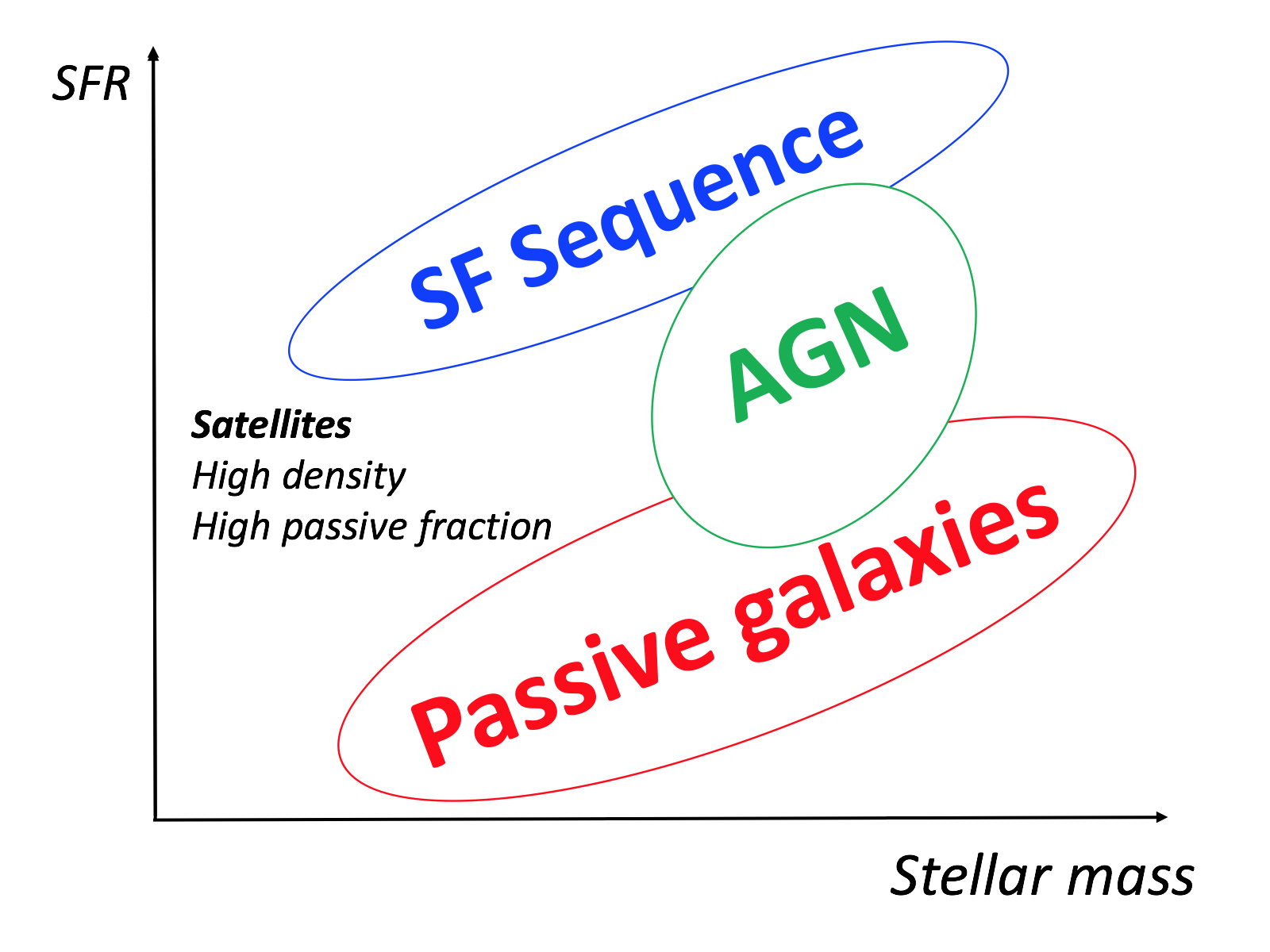}
\centering
\includegraphics[width=1\columnwidth]{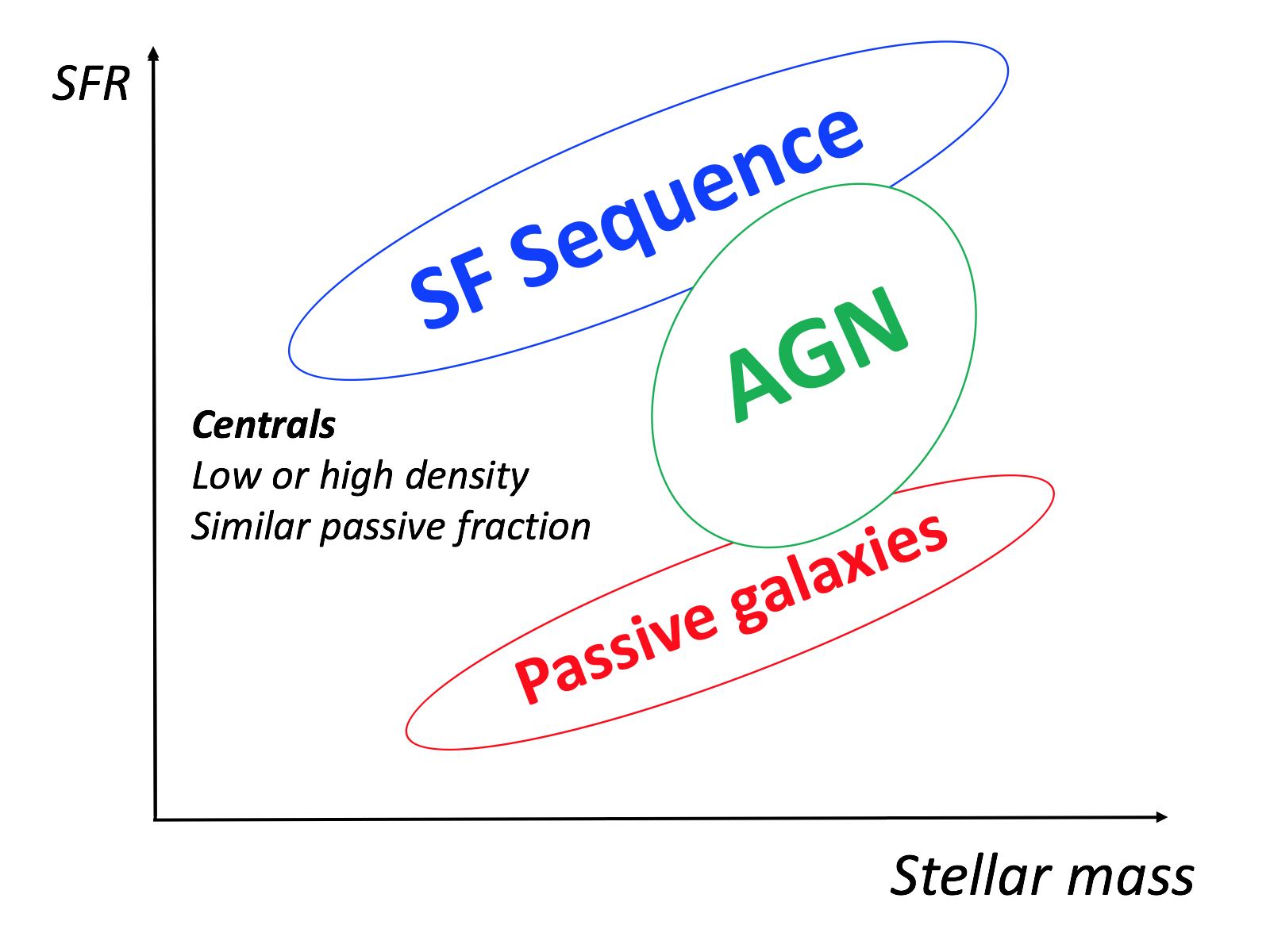}
\caption{Cartoons as illustrations of the distributions of satellites (top panels) and centrals (bottom panel) in SFR$-$M$_*$ diagram. For satellite galaxies, there is a higher passive fraction in higher density regions (right) than lower density regions (left). The strong correlation between passive fraction and environment could lead to significant bias in the AGN fraction defined by all galaxies. For central galaxies, the passive fractions are similar in low or high density regions, in which case the definitions of AGN fraction will lead to the same results. \label{figure 7}}
\end{figure*}

The group identity alone could only provide limited evidence for our scenario, it is necessary to examine the environmental dependence by using more indicators such as overdensity. To do this, we first start from satellite galaxies which are known to be susceptible to local environment. In the top-left panel of Figure \ref{figure 5}, we could see clearly that the AGN fraction defined by all galaxies decreases with increasing overdensity at $\rm log(1+\delta) > 0.8$, which even goes down to a half in satellites above $\rm 10^{10.5}M_{\odot}$. This result, together with the dichotomy in Figure \ref{figure 4}, may suggest that AGNs prefer underdense regions, especially for those with massive host galaxies. In fact, similar argument has been addressed by many work (e.g. \citealt{Kauf 04, Silver 09, Sab 13, Col 17}). However, by using our new AGN fraction defined as the number of AGNs relative to the star-forming galaxies only, we find there is no significant dependence of AGN fraction on overdensity. There might only be a hint that AGN fraction slightly increases with overdensity at some given stellar mass (the top-right panel of Figure \ref{figure 5}). This demonstrates that AGN activity does not strongly depend on the environment of its host galaxy, which is apparently at odds with most of the results in literature.   

Bottom panels of Figure \ref{figure 5} show the AGN fraction of central galaxies as a function of overdensity. Unlike satellites, the AGN fraction of centrals show little dependence on environment by using either definition. We will interpret the discrepancy between centrals and satellites in section \ref{D}. Therefore, the basic conclusion is that AGNs in general (no matter in central or satellite galaxies) have no significant preference on environment. 

Furthermore, we investigate the dependence of AGN fraction on group halo mass. In Figure \ref{figure 6}, there is no significant dependence of AGN fraction on halo mass with only a slight preference in more massive halos. This again supports our scenario that AGN activity does not strongly depend on environment. Here we only consider satellites rather than centrals mainly for two particular reasons. First, satellites are presumably more susceptible to environmental effects (e.g. \citealt{van 08, Peng 12, Wet 12, Kov 14, Lin 14a, Kno 15, Dar 16, Hen 17, Str 19}). Second, the stellar mass of centrals are tightly correlated with its parent halo mass \citep{Peng 12}, which may lead to some entanglement between mass and environment.

\section{Discussions \label{D}}
In the investigation of environmental dependence of AGN activity, the different definitions of AGN fraction can lead to inconsistent conclusions. It is hence important to evaluate which definition can better represent the occurrence of AGN activities. The major difference between the two definitions (see equation \ref{eq_all} and \ref{eq_sf}) is whether the passive population is incorporated in the denominator: if we only use AGN and star-forming galaxies as the denominator instead of all galaxies, the significant dependence of AGN fraction on environment will disappear. We believe our new definition of AGN fraction defined as the number of AGNs relative to the number of star-forming galaxies can better reflect the intrinsic correlation between AGN and environment, which is supported by the following two reasons.

The first reason is that a vast number of passive galaxies were quenched at high redshift (e.g. $z>1$), especially massive galaxies. They could have been quenched by AGN and/or stellar feedback (or by any other physical mechanisms) in the high redshift Universe. For instance, some of the massive red galaxies are elliptical galaxies that have been already ``red nuggets'' before $z\sim2$ \citep{Hua 13a, Hua 13b}. These galaxies cannot help us understand the interrelationship between AGN and environment at $z\sim0$. We should hence exclude these fossil passive galaxies when describing the prevalence of AGNs in the local Universe. 

The second reason is the strong correlation between passive fraction and environment i.e. both the fraction of the passive galaxies and different morphological types of galaxies have been found strongly correlated with the local environment (e.g. \citealt{Dre 80, Kauf 04, Bal 06, Peng 10, Peng 12}). The correlation can be illustrated by the cartoons in Figure \ref{figure 7}: some fraction (depending on density) of passive satellite galaxies were quenched by external environmental effects including tidal stripping (e.g. \citealt{Rea 06}), ram-pressure stripping (e.g. \citealt{Gun 72, Aba 99, Qui 00}), harassment (e.g. \citealt{Far 81, Moo 98}) and strangulation (e.g. \citealt{Lar 80, Fel 10, Peng 15}), instead of internal effects which AGN feedback is assumed to be responsible for. For satellite galaxies, there is a higher passive fraction in higher density regions than lower density regions. The strong correlation between passive fraction and environment could lead to significant bias in the AGN fraction defined relative to all galaxies i.e. more passive galaxies in dense environment will dilute the frequency of AGN activity. 

For central galaxies, however, the AGN fraction show little dependence on environment at fixed stellar mass by using both definitions (Figure \ref{figure 5}). According to \citet{Peng 12}, centrals cannot be environmentally quenched, in other words, centrals can only be quenched by internal mass-related processes. Therefore, it is reasonable to expect that the passive fraction of centrals does not change with overdensity, in which case the choice of definitions make no difference (see Figure \ref{figure 7} for an illustration).

Our AGN fraction defined as the number of AGNs relative to the star-forming galaxies can eliminate the strong correlation between passive fraction and environment, thus manifesting the intrinsic correlation between environment and the occurrence of AGN activity. This is a rather novel definition of AGN fraction compared with the previous AGN fraction defined as the number of AGNs relative to all galaxies including passive and star forming ones. Coincidentally, we notice that in a most recent work, \citet{Ami 19} constructed two separate samples, AGN and star-forming galaxies (SFG), and found the AGN fraction (with respect to SFG) weakly depends on the distance from the brightest cluster member. This is consistent with our results and also supports our definition of AGN fraction.

\section{Conclusions \label{Con}}
In summary, we use the SDSS DR 7 sample from  \citet{Peng 10}, which contains 214,091 galaxies with reliable measurements of galaxy properties including stellar mass, SFR, over-density, in the redshift range of $0.02 < z < 0.085$. Group properties including central satellite classification and halo masses are taken from \citet{Yang 07} group catalog. Through the emission-line selection on BPT diagram, galaxies are classified as star-forming, AGN, composite, low S/N LINER, etc. \citep{Bri 04}. Our goal is to investigate the environmental dependence of AGN activity in local Universe, and the results are as follows:

(1) AGNs, composites, low S/N LINERs and non-AGN galaxies that have a similar stellar mass all have similar spatial distributions.

(2) At a given stellar mass, the sSFR distribution of AGN hosts does not depend on environment, while that of the normal non-AGN galaxies strongly depends on environment. AGNs are predominantly located in the transition area (green valley) between passive and star-forming populations, indicating that AGN may be associated with galaxy quenching process, as in previous studies. 

(3) In previous similar studies, the AGN fraction is commonly defined as the number of AGNs relative to all galaxies (including passive and star forming ones). With this definition, the AGN fraction does decrease with increasing over-density for satellite galaxies, and satellite galaxies have a lower AGN fraction than the centrals. As illustrated in Figure \ref{figure 7}, this apparent dependence of AGN fraction on environment is largely due to the fact that the fraction of passive galaxies strongly depends on environment. Therefore, in order to investigate the intrinsic correlation between AGN and environment, especially under the assumption that AGN feedback is responsible for star formation quenching, the AGN fraction should be defined as the number of AGNs relatively to the star-forming galaxies only. With this new definition of AGN fraction, we find little dependence of AGN activity on environment in terms of overdensity, central/satellite dichotomy and group halo mass.

(4) There is only marginal evidence that the optical-selected AGNs might prefer denser regions (Figure \ref{figure 5} and \ref{figure 6}), which may be due to more frequent interaction of galaxies or higher merger rate in groups. Many theoretical models believe that AGNs are largely triggered through the process of galaxy mergers or strong tidal interactions (e.g. \citealt{Kau 00, Hop 06, Hop 08}). Mergers can drive gas inflows to the galaxy nucleus, fueling starbursts and may trigger AGN activity (e.g. \citealt{Ram 12, Jian 12, Kam 13, Kav 15, Chi 15}). However, the connection between nuclear activity and galaxy mergers and interactions is not yet clear (see \citealt{Hec 14} for a review). 

Our new findings support the scenario that internal secular evolution is the predominant mechanism of triggering AGN activity. The weak positive correlation between AGN fraction and environment for both centrals and satellites suggests that external environment related processes only play a minor role. One possible mechanism being responsible for fueling AGNs whose host galaxies are not paired could be a slow inflow of gas driven by non-axisymmetric perturbations \citep{Kaw 06, Elli 11, Vil 12, Hop 14, Kou 14}. Such perturbations in the underlying mass distribution of a disk can be caused by bars, oval distortions and even spiral arms (e.g., \citealt{Kor 04}; \citealt{Ath 08}). In particular, bars may fuel AGNs by driving radial gas inflows, and  some previous studies have shown that the incidence of bar formation may not depend significantly on the local environment \citep{van 02, Li 09, Mar 11, Lee 12, Lin 14b}. Other studies found a higher bar fraction in dense environment like clusters \citep{Bar 09, Bar 11}, which seems to reconcile with our conclusion (4). Nevertheless, the relationship between bars and environment has been controversial, and it is also unclear whether bars can trigger AGNs in the center of the host galaxies. We should be therefore cautious with the interpretations of the results. 

In addition, the AGNs in our sample are optical-selected in the local Universe. It would be interesting to apply our approach to other AGN populations, such as the X-ray selected AGNs and radio selected AGNs, to test whether our conclusion still holds. With upcoming major facilities such as MOONS on VLT and PFS on Subaru, it will become possible to carry out SDSS-like surveys at higher redshifts. This will allow further investigation of the dependence of AGN activity on environment as a function of cosmic time, shedding light on the AGN triggering mechanism at different epochs. 

\section{Acknowledgement}
We are grateful to Simon Lilly and Luis Ho for the productive discussions and useful comments. We thank the anonymous referee for useful comments. This work is supported by the NSFC Grant No. 11773001 and National Key R\&D Program of China Grant 2016YFA0400702. K.G. acknowledges the support from the Beijing Natural Science Foundation (Youth program) under grant No.1184015. X.K. acknowledges the support by the National Key R\&D Program of China Grant 2015CB857004 and 2017YFA0402600, and the National Natural Science Foundation of China Grant No. 11320101002, No. 11421303, and No. 11433005.

 \end{document}